\documentclass[twocolumn,preprintnumbers,nofootinbib]{revtex4-1} 
\usepackage{latexsym} 
\usepackage{mathptm}
\usepackage{amsmath}
\usepackage{amssymb}
\usepackage{graphicx}
\def\beq{\begin{equation}}
\def\eeq{\end{equation}}
\def\beqn{\begin{eqnarray}}
\def\eeqn{\end{eqnarray}}
\begin{document}
 
\title{Bounds on an energy-dependent and observer-independent speed of light from violations of locality}
\author{Sabine Hossenfelder} 
\affiliation{NORDITA, Roslagstullsbacken 23, 106 91 Stockholm, Sweden}

\begin{abstract}
We show that models with deformations of special relativity that
have an energy-dependent speed of light have non-local effects. The requirement that the arising
non-locality is not in conflict with known particle physics allows us to derive strong bounds on 
deformations of special relativity and rule out a modification to first order in energy over the Planck mass.
\end{abstract}
\maketitle

Modifications of special relativity have recently obtained increased
attention since measurements of gamma ray bursts observed by the Fermi Space Telescope have
now reached a precision high enough to test an energy-dependence
of the speed of light to first order in the photon's energy over the Planck mass 
\cite{Science,Nature,AmelinoCamelia:2009pg}. While such modifications could also be caused by an actual breaking of Lorentz-invariance
that introduces a preferred frame, Lorentz-invariance breaking is
subject to many other constraints already \cite{Maccione:2007yc}. This makes deformations
of special relativity ({\sc DSR}) \cite{AmelinoCamelia:2000mn,AmelinoCamelia:2000ge,KowalskiGlikman:2001gp,AmelinoCamelia:2002wr,Magueijo:2002am},
that preserve observer-independence and do not introduce a preferred frame, the prime 
candidate for an energy-dependent speed of light. We will here show 
that {\sc DSR} necessitates violations of locality that put a 
bound on an energy-dependent speed of light that is 23 orders of magnitude stronger 
than the recent 
measurements of gamma ray bursts. We will use units in which $c=\hbar=1$.
 
{\sc DSR} is motivated by finding modified Lorentz-transformations that allow
the energy associated to the Planck mass, $m_{\rm Pl}$, to remain invariant under
action of the transformations. 
In the cases of {\sc DSR} we will examine, the speed of light is a function of 
energy $\tilde c(E)$, such that this function is the same for all observers. 
Thus, in a different restframe
where $E$ was transformed into $E'$ under the deformed Lorentz-transformation, the
speed of light would be $\tilde c'(E') = \tilde c(E')$. This invariance 
of the functional behavior of the now energy-dependent speed of light is the key point of 
{\sc DSR}, and the modified transformations 
are constructed such that the function $\tilde c(E)$ can indeed remain invariant under a
change of reference frame, which would not be the case under ordinary Lorentz-transformations.

The modified transformations are commonly derived by requiring the invariance of
a modified dispersion relation \cite{AmelinoCamelia:2000mn} from which the speed of light 
can be obtained. These transformations were originally considered for momentum space. However, the
claim that {\sc DSR} makes predictions for the propagation of photons from gamma ray bursts 
clearly employs the energy-dependence of the speed of light in position space. We will in the following
show that the requirement of an energy-dependent {\sl and} observer-independent 
speed of light in position space results in an observer-dependence of what it means for two
events to be at the same point in space and time. This results in a violation of locality
in the sense that two observers in relative motion to each other cannot agree on
whether two events at the same point, and this disagreement is macroscopic
even for moderate relative velocities resulting in an inconsistent definition of space-time
location that is in conflict with already established physics. It should be stressed that this problem does
not occur in theories with an energy-dependent speed of light that actually break 
Lorentz-invariance. In this case the functional form of $\tilde c(E)$ will not remain
invariant under a change of reference frame.
 
Consider a
gamma ray burst ({\sc GRB}) at distance $L \approx 4$~Gpc that, for simplicity, has no motion
relative to a laboratory where it is detected. This source emits a
photon with $E_\gamma \approx 10$~GeV which arrives in the lab
restframe at $(0,0)$ inside a detector. Together with the 10 GeV 
photon there is a low energetic reference photon emitted. The energy of that photon can be as
low as wanted. 

In the {\sc DSR}-scenario we are considering the phase velocity 
depends on the photons' energy. To first order
\beqn
\tilde c(E) \approx \left( 1 + \alpha \frac{E}{m_{\rm Pl}} \right) + {\cal O} \left(\frac{E^2}{m^2_{\rm Pl}} \right) ~,
\label{alpha}
\label{cofe}
\eeqn
where we will neglect corrections of order higher than $E_\gamma/m_{\rm Pl}$ in the following. 
The important point is that Eq. (\ref{cofe}) is supposed to be observer-independent, such that it has the same form in every reference frame. This then requires the non-linear, deformed 
Lorentz-transformations in momentum space. Four our purposes it is sufficient to know that the Lorentz-transformations receive to lowest order a correction in $E/m_{\rm Pl}$.
To ease the discussion
we consider the case $\alpha < 0$ such that the speed of light decreases with increasing energy. The argument however does not depend on the sign.  

The 
higher energetic photon is slowed down and arrives
later than the lower energetic one. One has for the difference $\Delta T$ 
between the arrival times of the high and low energetic photon 
\beqn
\Delta T = L \left(\frac{1}{\tilde c (E_\gamma)} - 1 \right) = L \frac{E_\gamma}{m_{{\rm Pl}}} + {\cal O} \left(\frac{E_\gamma^2}{m^2_{\rm Pl}} \right) ~. \label{DeltaT}
\eeqn
With 4 Gpc $\approx 10^{26}$~m, $E_\gamma \approx 10^{-18} m_{{\rm Pl}}$, the delay is of the order 1 second. Strictly speaking, we should take into account the cosmological redshift since the photon propagates
in a time-dependent background. However, for the purpose of estimating the effect it will suffice to 
consider a static background, since using the proper general relativistic expression does not change the
result by more than an order of magnitude \cite{Ellis:2002in,Jacob:2008bw}.


\begin{figure}[ht]
\includegraphics[width=7cm]{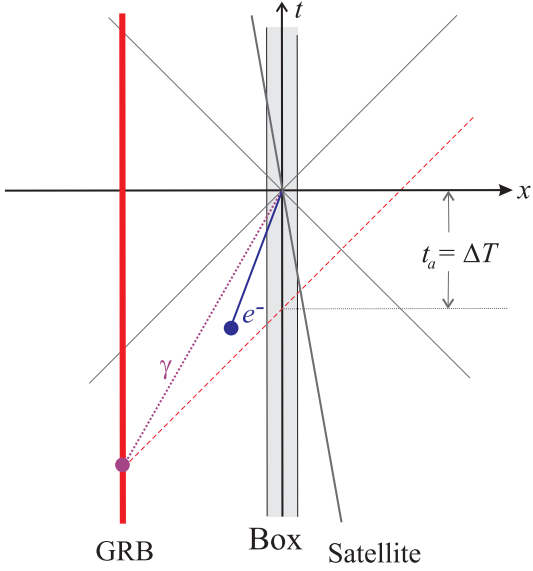}
\caption{Earth frame. The {\sc GRB} (thick solid) emits at the same time a low energetic photon (thin, long dashed) and
a highly energetic, slowed-down photon (dotted). The slowed-down photon crosses the worldline of an electron (solid) 
that was emitted nearby the detector (shaded area). A satellite flies by towards the {\sc GRB}.}
\label{1}
\end{figure}


We further consider an electron at $E_e \approx 10$~MeV emitted from a source in
the detector's vicinity such that it arrives together with the high-energetic photon at $(0,0)$ inside
the detector. The low-energetic photon leaves the {\sc GRB} together with the high-energetic photon 
at $(x_e, t_e) = (-L, - L/\tilde c(E_\gamma)$. It arrives in the detector at $(x_a, t_a)=(0,L(1-1/\tilde c(E_\gamma))$,
by $-t_a$ earlier than the electron.
We have chosen the emission time such that $-t_a=\Delta T$ in the lab frame and the electron arrives
with the same delay after the low energetic photon as the high energetic photon. With an energy of 10~MeV, the
electron is relativistic already, but any possible energy-dependent {\sc DSR} effect is at least 3
orders of magnitude smaller than that of the photon, and due to the electron's nearby emission the effects cannot accumulate over a long distance. The electron's velocity is $v_e \approx   1 - 10^{-3}$.

Inside the detector at $x=0$ the photon scatters off the electron. The photon changes 
the momentum of the electron, which triggers a bomb and the lab blows up. 
That is of course completely irrelevant. It only matters that the elementary scattering 
process can cause an irreversible and macroscopic change. This setup is depicted in Figure \ref{1}.

Now let us consider a team of physicists in a satellite moving towards the {\sc GRB} who observe and try to describe the 
processes in the lab. We will denote the coordinates of that restframe with $(x',t')$. 
 The satellite crosses the lab just when the bomb blows off at $(0,0)$. The typical speed of a 
satellite relative to Earth-bound laboratories is $v_S= -10$ km/s, and $\gamma_S \approx 1+10^{-9}$ for the relative motion between lab and satellite.   In the satellite frame, the electron's
and photons' energies are blueshifted. We have
\beqn
E'_\gamma &=& {\sqrt\frac{1-v_S}{1+v_S}} E_\gamma + {\cal O} \left(\frac{E^2_\gamma}{m^2_{{\rm Pl}}} \right)  ~,
\eeqn
and the energy of the very low energetic photon remains very low energetic.
In the satellite frame the time passing between the arrival of the low energetic
reference photon at the satellite and the electron at $x'=0$ is
\beqn
t'_a = \frac{L}{\gamma_S} \frac{1/\tilde c(E_\gamma) -1}{1 - v_S}  ~. 
\eeqn

\begin{figure}
\includegraphics[width=7cm]{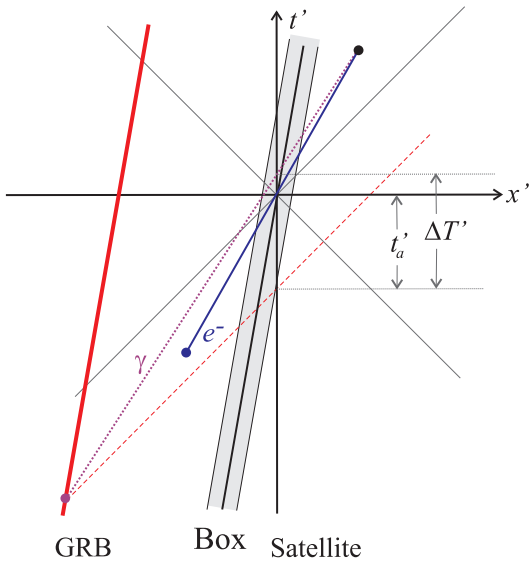}
\caption{Satellite frame. Same as in Figure \ref{1}, as seen from the satellite frame. The highly energetic photon is
slowed down further and does not meet the electron in the detector.}
\label{2}
\end{figure}

With higher energy, the speed of the electron increases. The speed of the 
photon also changes but, and here is
the problem,  according to {\sc DSR} 
the function $\tilde c$ is {\sl observer-independent}. In the satellite
frame one then has

\beqn
\tilde c (E_\gamma') = 1 - \frac{E'_\gamma}{m_{{\rm Pl}}} = 1 - {\sqrt\frac{1-v_S}{1+v_S}} \frac{E_\gamma}{m_{\rm Pl}} + {\cal O} \left(\frac{E_\gamma^2}{m^2_{\rm Pl}} \right)~,
\eeqn
and the distance the photons travel is 
$L'= \gamma_S \left( v_S/\tilde c(E_\gamma) - 1 \right) L$.
Thus, the time passing between the arrival of the low-energetic and the 
high-energetic photon in the satellite is
\beqn
\Delta T' = \frac{E'}{m_{\rm Pl}} L' = \frac{1-v_S}{1+v_S} \Delta T +  {\cal O} \left(\frac{E_\gamma^2}{m^2_{{\rm Pl}}} \right)~. 
\label{dt}
\eeqn
The question arises whether there could also be some energy-dependence in the transformation of $L$ in position-space. We will discuss this possibility later.
With the above, in the satellite frame the high energetic photon thus arrives later than the electron by 
\beqn
 \Delta T' - t'_a  = \left( \frac{1-v_S}{1+v_S} - \frac{1}{ \gamma_S(1-v_S)} \right)  \Delta T +  {\cal O} \left(\frac{E_\gamma^2}{m^2_{\rm Pl}} \right)  ~.
\eeqn
Inserting $(1/\gamma_S) \approx 1 - (v^2_S)/2$ for $v_S\ll 1$, one finds
\beqn
\Delta T'  - t'_a \approx -3 \Delta T \left(v_S - v_S^2/2 \right)  \approx  10^{-5} \Delta T ~.
\eeqn
In the satellite frame, depicted in Figure \ref{2}, 
the high energetic photon thus misses the electron by $\approx 10^{-5}$ seconds, and is still lagging behind as much as a kilometer when it arrives in the detector. The photon then cannot scatter off the electron in the detector, and
the electron cannot trigger the bomb to blow up the
lab. The physicists in the satellite are puzzled.


An assumption we implicitly made for this derivation was 
that the quantum mechanical space- and time-uncertainties $\Delta t, \Delta x$ are not modified in {\sc DSR},
such that the GeV photon can be considered peaked to a $\Delta t$ smaller than the distance to the electron at arrival.

Whether or not the wave function spreads in {\sc DSR} depends on the
interpretation of the modified dispersion relation. It is supposed to describe the
propagation of a particle in a background with quantum gravitational
effects. Yet the question is whether this modification should be understood as
one for a plane wave or for a localized superposition of plane waves already. In
the first case a wave-packet would experience enhanced dispersion, in the latter
case not. In the absence of a derivation, both seems plausible, so let us just examine the 
possibilities. There either
is a modification, or there is not. The above covered the case without modification.

In case there is a modification caused
by a dispersion of the wave-packet, then the uncertainty of the slowed down, high energetic photon at arrival would 
be vastly larger than the maximal localization of the Heisenberg limit allows. If one starts with a Gaussian
wave-packet localized to a width of $\sigma_0$ at emission and tracks its spread with the modified dispersion
relation, one finds that to first order the now time-dependent width is
\beqn
\sigma(t) = \sigma_0 \sqrt{1 + \left( \frac{2 t}{m_{\rm Pl} \sigma^2_0}\right)^2}~.
\eeqn 
If we start with a width of $\sigma_0 \approx 1/E_\gamma$, then for $t \gg m_{\rm Pl} \sigma_0^2$ 
one has
$\sigma(t) \approx 2 t E_\gamma/m_{\rm Pl}$. Or, in other words,  
in the worst case the uncertainty of the wave-packet at arrival was about the same size as
the time delay $\Delta t \approx \Delta T$ and the photon 
would at arrival be smeared out over some hundred thousand
kilometers. A delay of $\Delta T$ with an uncertainty of $\Delta T$ is hard to detect, but it would also be  
impossible to find out whether or not the center of the wave-packet had been dislocated by a factor five orders of
magnitude smaller than the width of the wave-packet. 

However, the problem was caused by the unusual transformation behavior of $\Delta T$. To
entirely hide this behavior, the quantum mechanical uncertainty $\Delta t$ needed to be much larger than the delay 
$\Delta T - t_a$ in all restframes, such that it was practically unfeasible to ever detect a tiny difference
in probability with the photons we can receive, say, in the lifetime of the universe. 
Therefore, let us boost into a reference frame with $v=1-\epsilon$, such that $\gamma \approx 1/\sqrt{2 \epsilon}$.
The inequality that needs to be fulfilled to hide the delay $|\Delta T' - t_a' | \ll |\Delta t'|$ is then equivalent to 
$ |\epsilon - \sqrt{2/\epsilon}| \ll \epsilon$, which is clearly violated without even requiring extreme boosts.
To put in some numbers, for an observer in rest with the electron one has $\epsilon = 10^{-3}$, $\gamma \approx 20$ and $| \Delta T' -t_a' | \approx 10^4 \Delta t'$. 

Similarly, for $v=-1+\epsilon$, the requirement to hide the delay
takes the form
$ |2/\epsilon - \sqrt{\epsilon/ 2}| \ll 2/\epsilon$, which is also clearly violated. Though in this case the delay does not actually get much larger than the uncertainty, they
 both approach the same value. We would then be comparing the probability of interaction at
the center of the wave-packet with one at a distance comparable to its width. This would require several
photons to get a proper statistic, but it is a difference in probability that is feasible to measure within
the lifetime of the universe, and thus is still in conflict with observer-independence. Let us 
point out that we have considered here 
a photon whose approximate momentum uncertainty at emission is comparable to the mean value, which is quite large
already. If the
photon's momentum had instead a smaller uncertainty, e.g. $\approx 100$~MeV only, then the mismatch in timescales
was by two orders of magnitude larger.

To keep track of our assumptions, we used normal Lorentz-boosts to calculate the time span $t'_a$. 
This is based on already available data since the transformation behavior under 
Lorentz-boosts has been tested to high precision in particle physics experiments \cite{Reinhardt:2007zz}. For the time-dilatation in
particular, the decay-time of muons is known 
to transform as $\Delta t' = \gamma \Delta t$ up to a $\gamma$-factor of 30 to a precision 
of one per mille \cite{Bailey:1977de}.

The logic of the here presented argument is as follows. If there was a delay of the order 1 second for
the 10 GeV photon caused by an energy-dependent speed of light, then the requirement of 
observer-independence results in violations of locality that are in conflict with experiment already.
Note that it is not necessary to actually perform the above sketched experiment in all reference frames, since observer-independence
allows us to use cross-sections measured in our laboratories. This means in turn one can now use the 
knowledge of {\sc QED} processes combined with the measurements of Lorentz-boosts 
to constrain the possibility of there
being such a {\sc DSR} modification by requiring the resulting mismatch in arrival
times not to result in any conflict with existing measurements. 

The distance $L =$ some Gpc is as high as we can plausibly get in
our universe, and the $10$ GeV photon is as high as we have reliable observational
data from particles traveling that far. The center of mass energy of the electron
and the high energetic photon is $\sqrt{s} \approx 15$~MeV. The process
thus probes distances of $\approx 10$~fm. If the photon and
the electron were closer already than the distance their 
scattering process probes, we would not have a problem. 
Requiring $|\Delta T' - t'_a| < 10$~fm for boosts up to $\gamma = 30$ leads to a bound on the delay between the low
and high energetic photon of
\beqn
\Delta T < 10^{-23} {\rm s} \quad,
\eeqn 
or, if we reinsert $\alpha$ from Eq. (\ref{alpha}), $|\alpha| < 10^{-23}$. Note that this covers both cases, the one with and without
spread of the wave-packet. 

We have here not discussed all possible constraints that one could consider, for example different
scattering processes and their exact precision. We see now that this is not necessary, since the ratio $E_\gamma/m_{\rm Pl}$ is $\approx 10^{-18}$.
With the above constraint, we are thus already in the regime where second order modifications would become important. The here used
analysis however made use of the scaling in Eq. (\ref{dt}) and thus does not in this way apply to the second order modifications.
We can conclude however that present-day observations do already rule out a 
modification in the speed of light to first order in the energy over Planck mass.

To retrace our steps, the problem stems from the transformation behavior 
of $\Delta T$ in Eq. (\ref{dt}). This behavior is
a direct consequence of requiring the energy-dependent speed of light $\tilde c$ to be
observer-independent, together with applying a normal, passive, Lorentz-transformation
to convert the distance $L$ into the satellite restframe.

The formulation of {\sc DSR} in position space has been under debate.
It has been argued that the space-time metric and also the Lorentz-transformations in
position space should become energy-dependent 
\cite{Magueijo:2002xx,Kimberly:2003hp,Galan:2004st,Amelino-Camelia:2005ne,Hossenfelder:2006rr}. Now 
if one would use a 
modified transformation also on the coordinates, a transformation
depending on the energy of the photon, then $\Delta T$ might transform properly 
and both particles would meet also in the satellite frame. This would require that
the transformation of the distance $L$ was modified such that it converted
the troublesome transformation behavior of $\Delta T$ back into a normal 
Lorentz-transformation.   

This would imply that the distance between any two objects would depend on the energy of a photon
that happened to propagate between them. The distance between
the {\sc GRB} and the detector was then energy-dependent such that it got shortened in the
right amount to allow the slower photon to arrive in time together with the electron. 
That however would mean that the speed of the photon would not depend on its energy when
expressed in our usual low-energetic and energy-independent coordinates. 
The confusion here stems from having defined a speed from the dispersion relation 
without that speed a priori having any meaning in position space. This
possibility thus just reaffirms that
observer-independence requires the speed of light to be constant.

That {\sc DSR} implies a frame-dependent meaning of what
is ``near'' was mentioned already in
\cite{AmelinoCamelia:2002vy}. Serious conceptual problems arising from
this were pointed out in \cite{Schutzhold:2003yp,Hossenfelder:2006rr},
and here we have demonstrated a conflict with experiment to very
high precision. If {\sc DSR} was indeed the origin of time-delays of
highly energetic photons from {\sc GRB}s, then it would also
lead to macroscopic effects we would long have observed. Consequently,
{\sc DSR} cannot be cause of observable effects in {\sc GRB} spectra. 
 
{\sc DSR} is motivated by the idea that the Planck energy
should be observer-independent which then leads to deformed Lorentz-transformations
and an energy-dependent speed of light. We have here seen that such an 
energy-dependent speed of light that is also observer-independent 
implies violations of locality that are strongly constrained by
experiment. It has however been argued in \cite{Hossenfelder:2006cw} that
the requirement of the Planck scale being observer-independent does not
necessitate it to be an invariant of Lorentz-boosts, since the result of
such a boost does not itself constitute an observation. It is sufficient 
that experiments made are in agreement over that scale. In
particular if the Planck length plays the role of a fundamentally minimal
length no process should be able to resolve shorter distances. This does
require a modification of interactions in quantum field theory at
very high center-of-mass energies and small impact parameters, but it
does not necessitate a modification of Lorentz-boosts for free particles. 
In these models the speed of light remains constant.

An extended version of the here presented argument and further discussion can be
found in \cite{Hossenfelder:2009mu}.
\section*{Acknowledgements}

I want to thank Giovanni Amelino-Camelia, Stefan Scherer, and Lee Smolin for helpful
comments.


{\small
\begin{thebibliography}{99}



\bibitem{Science} The Fermi LAT and Fermi GBM Collaborations, 
Science {\bf 323} (2009) 1688. 

\bibitem{Nature}
A.~A.~Abdo {\sl et al}, 
Nature {\bf 462} (2009) 331.

\bibitem{AmelinoCamelia:2009pg}
  G.~Amelino-Camelia and L.~Smolin,
  Phys.\ Rev.\  D {\bf 80}, 084017 (2009)
  [arXiv:0906.3731 [astro-ph.HE]].


\bibitem{Maccione:2007yc}
  L.~Maccione, S.~Liberati, A.~Celotti and J.~G.~Kirk,
  JCAP {\bf 0710}, 013 (2007)  [arXiv:0707.2673 [astro-ph]].

\bibitem{AmelinoCamelia:2000mn}
  G.~Amelino-Camelia,
  Int.\ J.\ Mod.\ Phys.\  D {\bf 11}, 35 (2002)
  [arXiv:gr-qc/0012051].

\bibitem{AmelinoCamelia:2000ge}
  G.~Amelino-Camelia,
  Phys.\ Lett.\  B {\bf 510}, 255 (2001)
  [arXiv:hep-th/0012238].

\bibitem{KowalskiGlikman:2001gp}
  J.~Kowalski-Glikman,
  Phys.\ Lett.\  A {\bf 286}, 391 (2001)
  [arXiv:hep-th/0102098].

\bibitem{AmelinoCamelia:2002wr}
  G.~Amelino-Camelia,
  Nature {\bf 418}, 34 (2002)
  [arXiv:gr-qc/0207049].

\bibitem{Magueijo:2002am}
  J.~Magueijo and L.~Smolin,
  Phys.\ Rev.\  D {\bf 67}, 044017 (2003)
  [arXiv:gr-qc/0207085].






\bibitem{Ellis:2002in}
  J.~R.~Ellis, N.~E.~Mavromatos, D.~V.~Nanopoulos and A.~S.~Sakharov,
  Astron.\ Astrophys.\  {\bf 402}, 409 (2003)
  [arXiv:astro-ph/0210124].

\bibitem{Jacob:2008bw}
  U.~Jacob and T.~Piran,
  JCAP {\bf 0801}, 031 (2008)
  [arXiv:0712.2170 [astro-ph]].


\bibitem{Reinhardt:2007zz}
  S.~Reinhardt {\it et al.},
  Nature Phys.\  {\bf 3} (2007) 861.

\bibitem{Bailey:1977de}
  J.~Bailey {\it et al.},
  Nature {\bf 268} (1977) 301.
 

\bibitem{Magueijo:2002xx} 
J.~Magueijo and L.~Smolin, 
Class.\ Quant.\ Grav.\ {\bf 21}, 1725 (2004) [arXiv:gr-qc/0305055]. 

\bibitem{Kimberly:2003hp} 
D.~Kimberly, J.~Magueijo and J.~Medeiros, 
Phys.\ Rev.\ D {\bf 70}, 084007 (2004) [arXiv:gr-qc/0303067]. 
 
\bibitem{Galan:2004st} 
P.~Galan and G.~A.~Mena Marugan, 
Phys.\ Rev.\ D {\bf 70}, 124003 (2004) [arXiv:gr-qc/0411089]. 

\bibitem{Amelino-Camelia:2005ne} 
G.~Amelino-Camelia, 
Int.\ J.\ Mod.\ Phys.\ D {\bf 14}, 2167 (2005) [arXiv:gr-qc/0506117]. 

\bibitem{Hossenfelder:2006rr}
  S.~Hossenfelder,
  Phys.\ Lett.\  B {\bf 649}, 310 (2007)
  [arXiv:gr-qc/0612167].

\bibitem{AmelinoCamelia:2002vy}
  G.~Amelino-Camelia,
  Int.\ J.\ Mod.\ Phys.\  D {\bf 11}, 1643 (2002)
  [arXiv:gr-qc/0210063].

\bibitem{Schutzhold:2003yp}
  R.~Schutzhold and W.~G.~Unruh,
  JETP Lett.\  {\bf 78}, 431 (2003)
  [arXiv:gr-qc/0308049].



\bibitem{Hossenfelder:2006cw}
  S.~Hossenfelder,
  Phys.\ Rev.\  D {\bf 73}, 105013 (2006)
  [arXiv:hep-th/0603032].

\bibitem{Hossenfelder:2009mu}
  S.~Hossenfelder,
  arXiv:0912.0090 [gr-qc].

\end{thebibliography}}
\end{document}